\documentclass[iop]{emulateapj}

\usepackage{color}
\newcommand{\Msun}{~M_\odot}

\newcommand{\ccm}{\rm ~cm^{-3}}
\newcommand{\kms}{\rm ~km~s^{-1}}
\newcommand{\ergs}{\rm ~erg~s^{-1}}

\newcommand{\wl}{\lambda}
\newcommand{\my}{~\mu}

\begin{document}

\title{Discovery of  molecular hydrogen  in SN 1987A}
\author{\phantom{}
Claes Fransson\altaffilmark{1}, Josefin Larsson\altaffilmark{2}, 
Jason Spyromilio\altaffilmark{3}, Bruno Leibundgut\altaffilmark{3}, Richard McCray\altaffilmark{4}, Anders Jerkstrand\altaffilmark{5}
}

\altaffiltext{1}{Department of Astronomy, The Oskar Klein Centre,
              Stockholm University, Alba Nova University Centre, SE-106 91 Stockholm, Sweden  
              }
\altaffiltext{2} {KTH, Department of Physics, and the Oskar Klein
Centre, AlbaNova, SE-106 91 Stockholm, Sweden}
\altaffiltext{3}  { European Southern Observatory, Karl-Schwarzschild-Strasse, 2, D-85748 Garching bei M\"unchen, Germany}
\altaffiltext{4}  { Department of Astronomy, University of California, Berkeley, CA 94720-3411, USA}
\altaffiltext{5}{Astrophysics Research Centre, School of Mathematics and Physics, Queen's University Belfast, Belfast, BT7 1NN, UK .}
\begin{abstract}
Both CO and SiO have been observed at early and late phases in SN 1987A. H$_2$ was  predicted to form at roughly the same time as these molecules, but was not  detected at early epochs. Here we report the detection of NIR lines from H$_2$ at $2.12 \ \mu$m and  $2.40 \ \mu$m in VLT/SINFONI spectra obtained between days 6489 and 10,120. The emission is concentrated to the core of the supernova in contrast to H$\alpha$ and approximately coincides with the [\ion{Si}{1}]/[\ion{Fe}{2}] emission detected previously in the ejecta.  Different excitation mechanisms and power sources of the emission are discussed. From the 
nearly constant H$_2$ luminosities we favour excitation resulting from the ${}^{44}$Ti decay.
\end{abstract}


\section{INTRODUCTION}
\label{sec-introd}
Observations of molecules in supernovae (SNe) have turned out to yield important diagnostics of the conditions in the cooling ejecta.  The first examples were discovered in SN 1987A shortly after explosion. CO was first detected 112 days after explosion \citep{Spyromilio1988} and persisted to over 600 days \citep{Bouchet1993}. Both the fundamental and overtone bands were detected. Somewhat later, at $\sim 160$ days,  the fundamental band of SiO was also detected \citep{Aitken1988,Roche1991}. From NLTE modeling of the rovibrational lines the masses were estimated to be $\sim 10^{-3} \Msun$ for CO \citep{Liu1992} and $\sim (4-8)\times10^{-4} \Msun$ for SiO \citep{Liu1994}. Recently at an age of 26 years both these molecules have  been detected in  rotational transitions at $\la 100$ K with ALMA \citep{Kamenetzky2013,Matsuura2015}. The CO mass was, however, at this epoch at least an order of magnitude larger than the estimate from the NIR observations in \cite{Liu1992}, indicating that most of the molecule formation may have occurred later than the last NIR observations.  

In addition to these molecules,  \citet[][hereafter CMC]{Culhane1995} predicted that  molecular hydrogen, H$_2$, would form  in the core. Their calculations showed that between 400 and 1000 days after explosion this would increase from a very low level to  $\sim 1\%$ of the total H abundance, where it would freeze out. \cite{Utrobin2005} later found that H$_2$ could be present  in the tenuous H envelope even at $\sim 2$ weeks after explosion, with an abundance of $\sim 10^{-4}$ of atomic H. 

Although predicted long ago, and constituting a substantial fraction of the ejecta mass, no detection of H$_2$ has been found in a SN. While H$_2$ has been detected in the Crab nebula \citep{Graham1990} the origin of this is unknown, although Graham et al. argue for the formation in the ejecta. In this paper we report clear evidence of H$_2$ from NIR rovibrational  ejecta lines  in SN 1987A observed with VLT/SINFONI at very late phases. 

\section{Observations}
\label{sec-mod}
  \begin{deluxetable}{lrcrc}
\tablecolumns{5}
\tablewidth{0pc}
\tablecaption{Log of H and K-band observations. 
\label{tab0}}
\tablehead{
\colhead{Date}&\colhead{Epoch} & \colhead{Band} & \colhead{Exposure}   & \colhead{Seeing}\\
\colhead{}&\colhead{days}&& \colhead{s}& \colhead{\arcsec}}
\startdata
2004-11-29&6489&K&1800&0.68\\
2005-10-30 -- 2005-11-14 &6832&K&5400&0.38\\
2005-10-22 -- 2005-11-18&6830&H&4200&0.31\\
2007-11-07 -- 2008-01-19&7615&K&9000&0.43\\
2007-11-07 -- 2008-01-09& 7602& H& 4800& 0.31\\
2010-11-05 -- 2011-01-02&8694&K&10800&0.35\\
2014-10-12 -- 2014-12-01 &10120&K&7200&0.58\\
\enddata
\end{deluxetable}
 The H and K-band observations,  covering $1.45-1.80 \mu$m and $1.95-2.45 \mu$m, respectively, were obtained 2004--2014 using the SINFONI Integral Field Spectrograph at the VLT \citep{Eisenhauer2003,Bonnet2004}. The log of the observations is given in Table \ref{tab0}.
The data were reduced using the standard ESO pipeline \citep{Modigliani2007}  and are discussed in detail in \cite{Kjaer2010}, together with modelling of the atomic emission. Additional dedicated software was developed by us to combine observations spread over many epochs. The absolute reproducibility (cross calibration) of the standard stars was better than 5\%. The encircled energy of the PSF for 80\% of the emission is $0.150 \times 0.125\arcsec$ in the 2005 K-band observation \citep{Kjaer2010}, which gives a good measure of the spatial resolution. The spectral resolution is $66 \kms$ and $110 \kms$ for the K- and H-band, respectively. 
 Because of the short exposure of the 2004 observations we do not discuss  this epoch further. The observations in 2004, 2005, 2011 and 2014 are discussed in \cite{Kjaer2007,Kjaer2010} and \cite{Larsson2013,Larsson2015}.
 
The Q-branch of H$_2$ at $\sim 2.4 \  \mu$m  is awkwardly located near the end of the K-band atmospheric window and also near the rise of the thermal background from the telescope, 
which could give rise to the appearance of features in the spectra. We have therefore examined individually all frames to ensure that the residual thermal and atmospheric backgrounds do not create artefacts that would mimic the Q-branch emission from the SN. 

\section{Results}
\label{sec-res}
\begin{figure}[!t]
\begin{center} 
\resizebox{\hsize}{!}{\includegraphics[angle=-0]{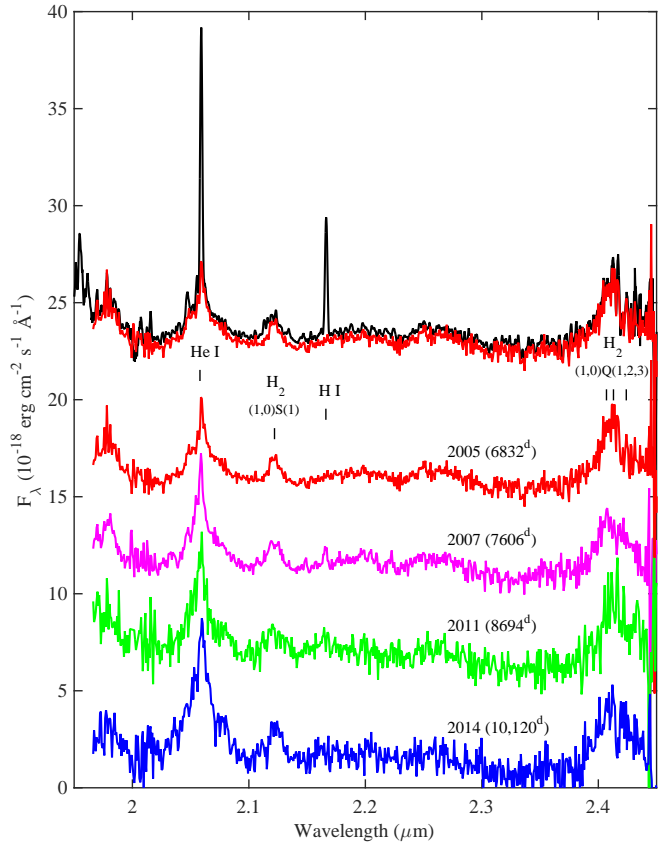}}
\caption{Compilation of K band ejecta spectra from 2005, 2007, 2011 and 2014 extracted from the elliptical region in Fig. \ref{fig2}. Each spectrum is shifted by $5\times 10^{-18} \rm erg ~ cm^{-2} ~ s^{-1} ~ \AA^{-1}$ relative to the previous. The upper  spectrum shows the 2005 spectrum before (black) and after subtraction of the scattered ring emission. Note the strong narrow components from this for the He I and Br$\gamma$ lines, while there are no such components in the H$_2$ lines. }
\label{fig1} 
\end{center}
\end{figure}

\begin{figure}[!t]
\begin{center}
\includegraphics[angle=-0,width=3.5cm]{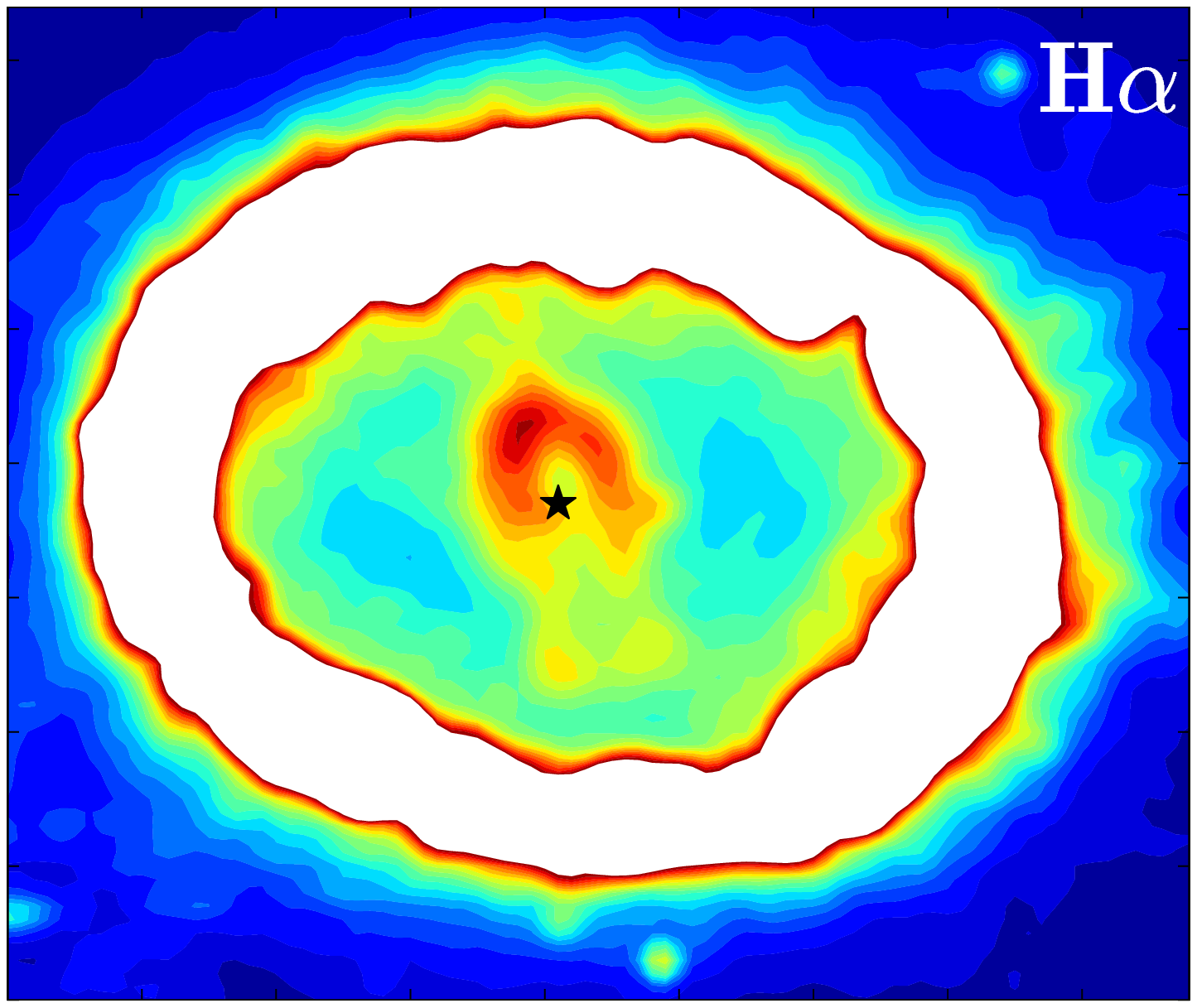}
\includegraphics[angle=-0,width=3.5cm]{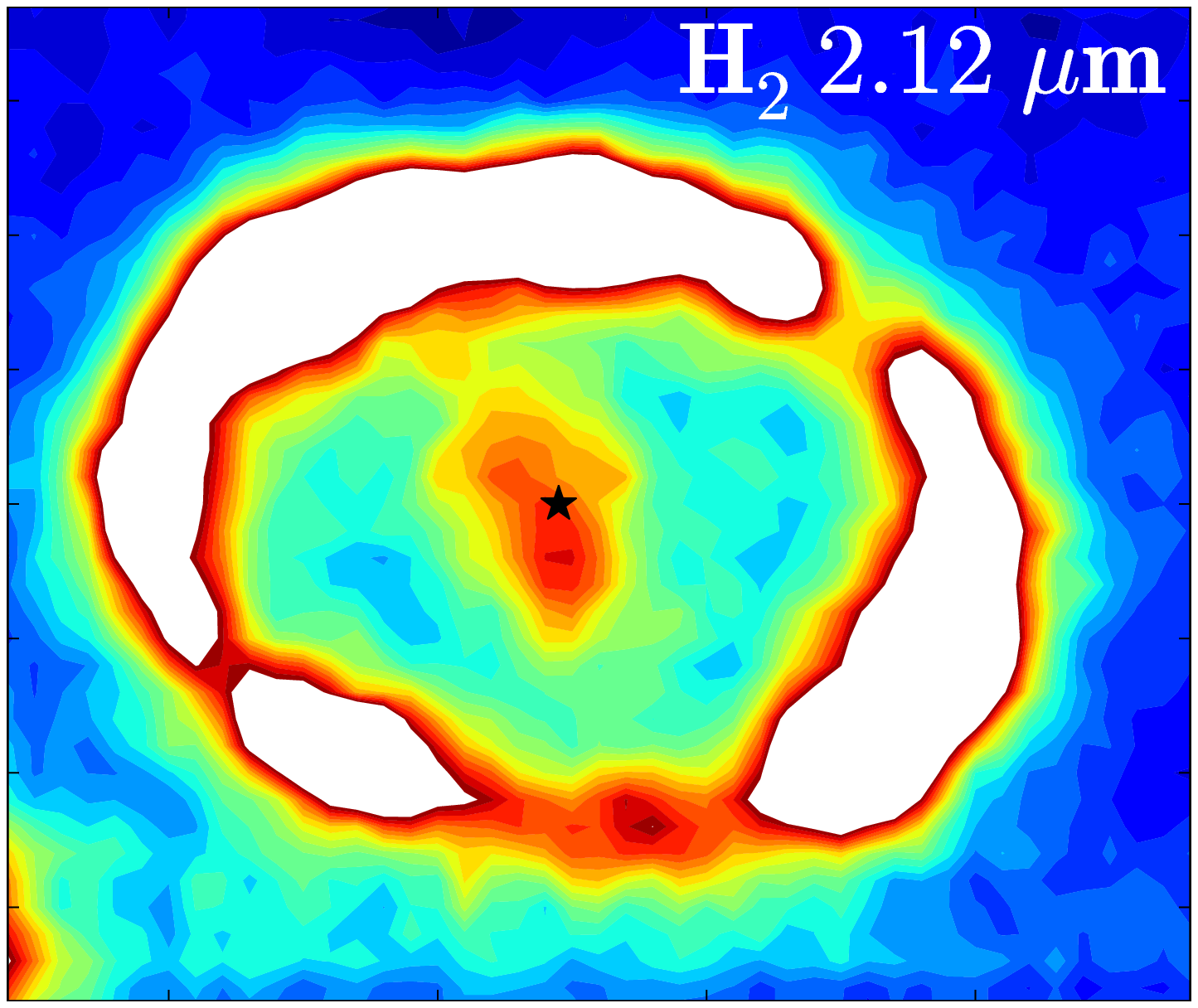}\\
\includegraphics[angle=-0,width=3.5cm]{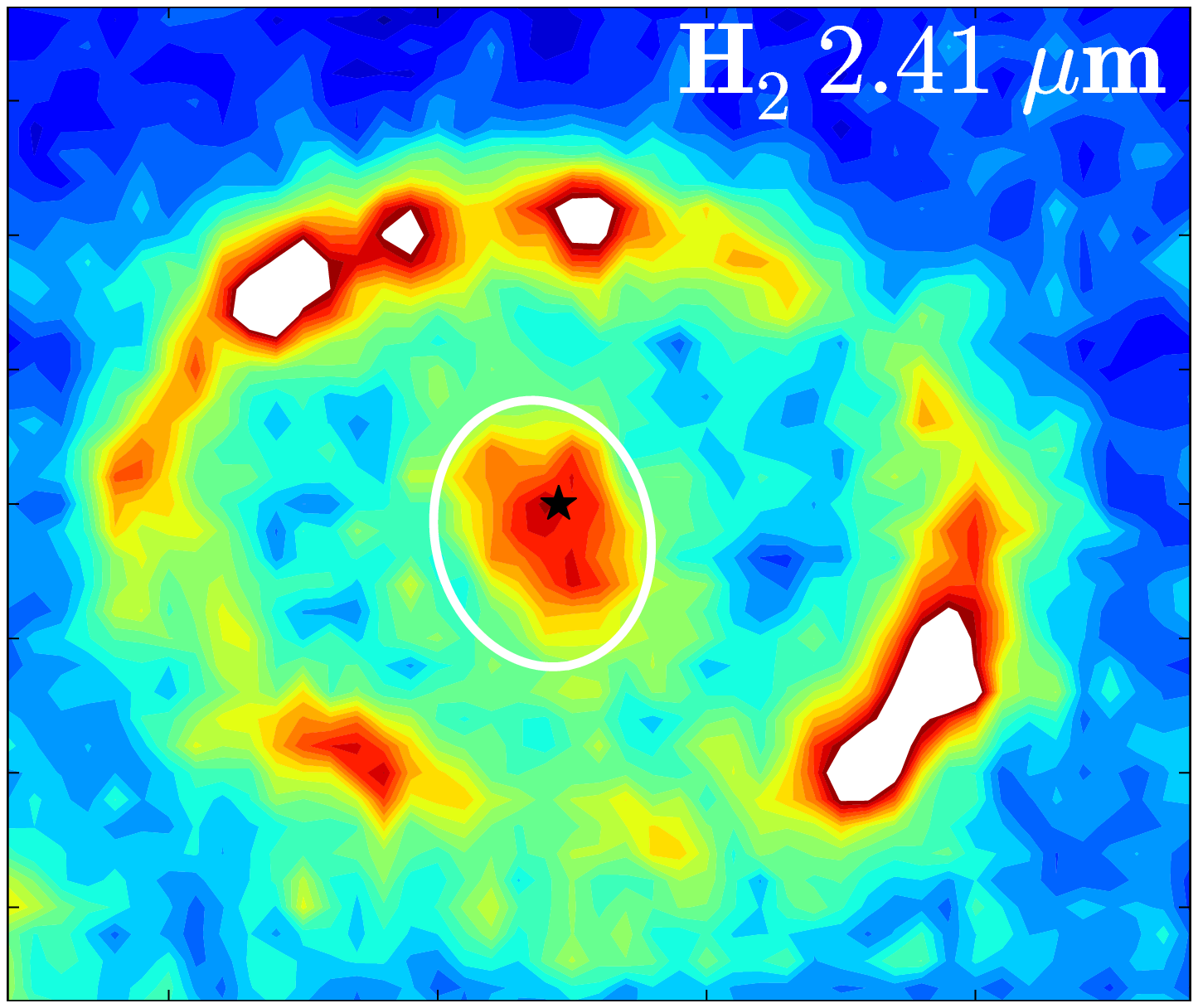}
\includegraphics[angle=-0,width=3.5cm]{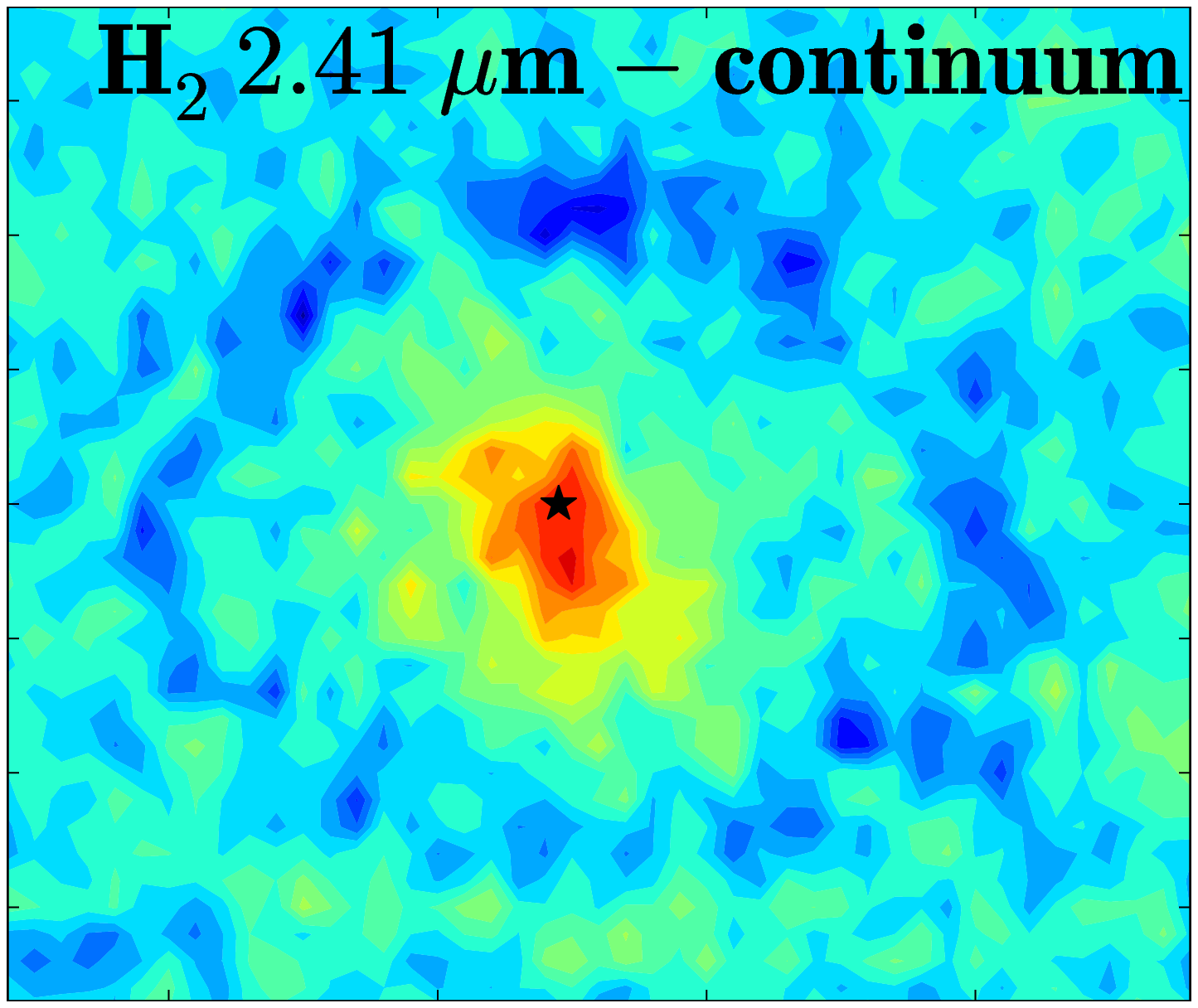}\\
\includegraphics[angle=-0,width=3.5cm]{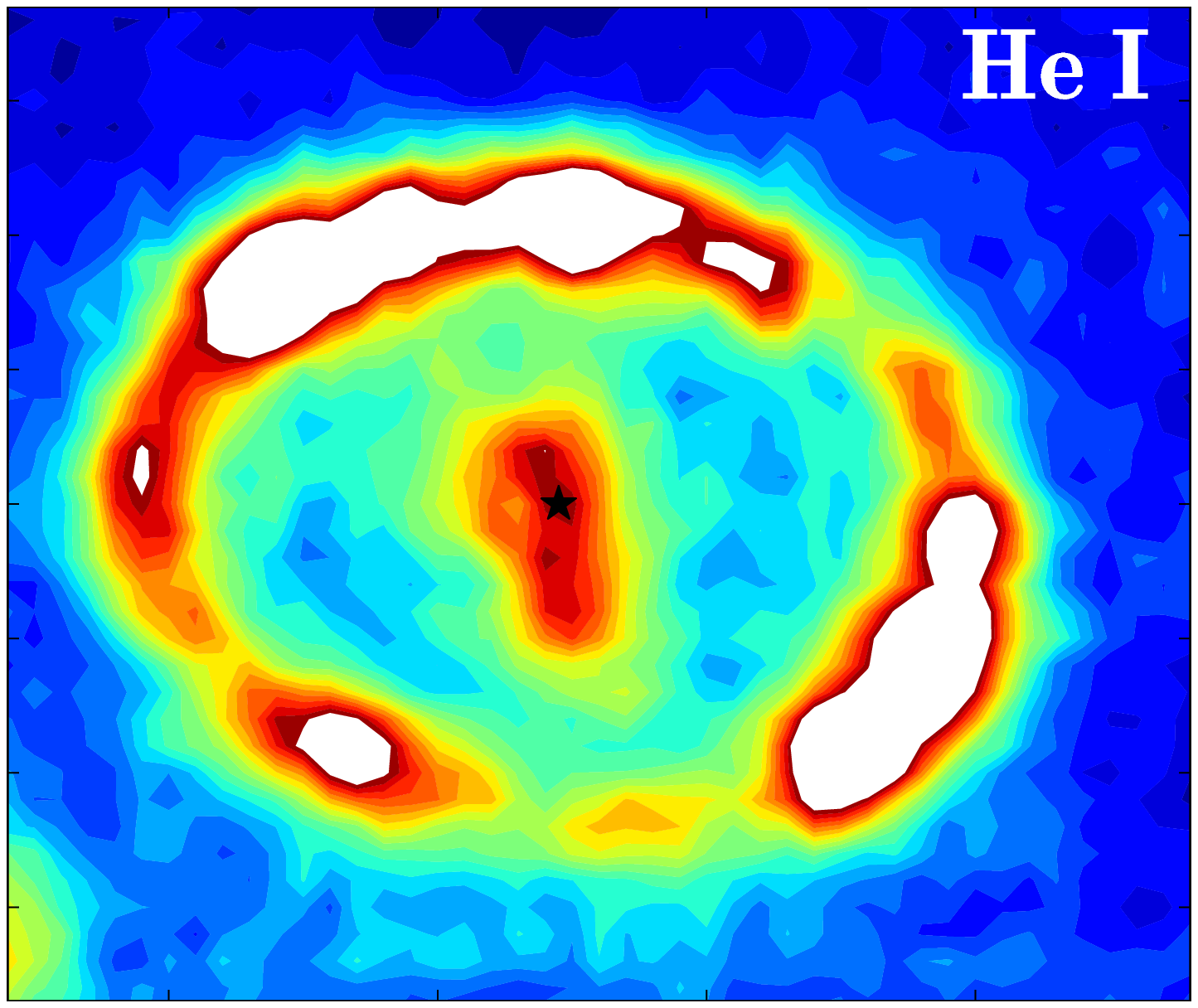}
\includegraphics[angle=-0,width=3.5cm]{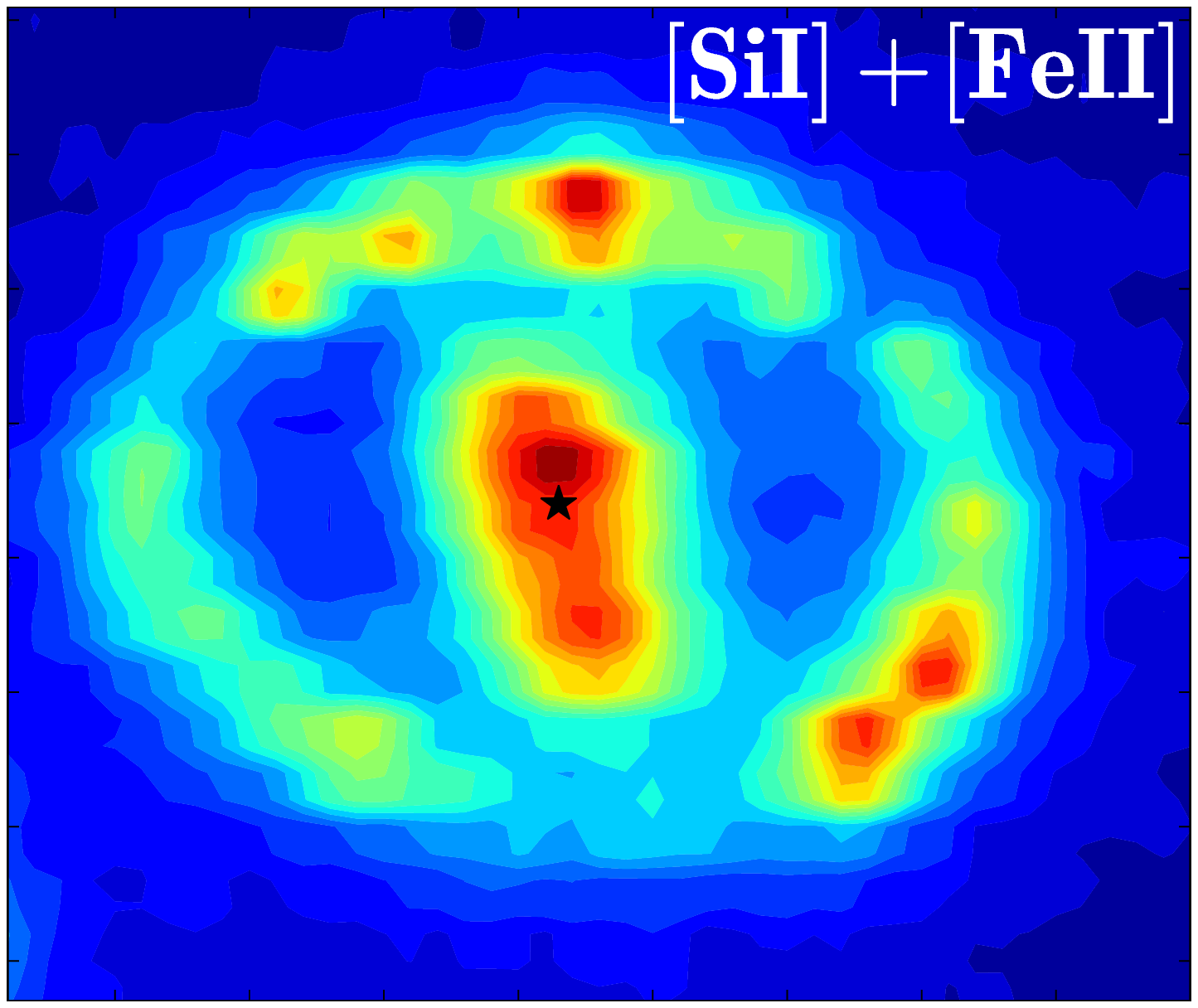}

\caption{Contour maps from 2005 of the spatial intensity distribution of H$\alpha$ (HST),  the H$_2$   2.12$\my$m, the H$_2$   2.40$\my$m band, the H$_2$   2.40$\my$m band without and with 'continuum' subtraction, \ion{He}{1} 2.058 $\my$m, and [\ion{Si}{1}]/[\ion{Fe}{2}] 1.644$\my$m (from SINFONI within $\pm 3000 \kms$ for each line). The major axis of the circumstellar ring is 1.6\arcsec \ and inclined by $\sim 43 \degr$ to the line-of-sight. The contours are linearly spaced between the maximum (red) and minimum (blue) ejecta region for each image. The star marks the center of the ring and the white ellipse in the H$_2$ panel the extraction region for the spectra in Fig. \ref{fig1}.  The weak extension to the north in the 2.12$\my$m image is a result of blending with the high velocity wing of the He I line.
}
\label{fig2} 
\end{center}
\end{figure}

In Fig. \ref{fig1} we show a compilation of the K-band  spectra  of the ejecta from 2005 to 2014,  days 6832,  7606, 8694 and 10,120. 
In order to minimise the contamination from the ring and  to compare the fluxes from the same co-moving  region we use an expanding elliptical aperture enclosing the emitting H$_2$ region, shown in Fig. \ref{fig2}. Scattered light from the ring was removed as in \cite{Larsson2015},  using the fact that lines from the ring are much narrower (FWHM $\sim 300 \kms$) compared to the ejecta lines (FWHM $\sim 2300 \kms$). This subtraction is illustrated in the upper spectrum in Fig. \ref{fig1}. The spectra were binned by a factor three. 

The strongest line  at all  epochs is the \ion{He}{1} 2.058 $\my$m line. We also see a faint Br$\gamma$ line at 2.166$\my$m.  In addition to these, there is a clear line at 2.12$\my$m and a strong, broad feature at 2.40$\my$m. Based on the absence of other natural candidates (see Sect. \ref{sec-mod}) we identify the 2.12$\my$m  line as the  (1,0) S(1) transition in H$_2$ and the feature at 2.40$\my$m as a blend of the (1,0) Q(1 -- 3) transitions at 2.406, 2.413, 2.423 $\my$m.  We do not detect the overtone CO band, which would peak between 2.30-2.35 $\mu$m.

In Fig.  \ref{fig2} we show the spatial distribution in 2005 of the 2.40$\my$m feature,  with and without subtraction of the 'continuum' in the line-free region between $2.300-2.345 \mu$m,  together with H$_2$ 2.12 $\mu$m, H$\alpha$ from HST \citep{Larsson2013},  and \ion{He}{1} 2.058 $\my$m, and [\ion{Si}{1}]/[\ion{Fe}{2}] from SINFONI \citep{Kjaer2010}.  The  emission from the circumstellar ring in the 2.12 $\my$m and 2.40$\my$m images is a result of H I line and continuum emission. 

From the figure we  note that the  2.12 $\my$m and 2.40$\my$m  emission have a markedly different distribution compared to H$\alpha$. Most of the emission is concentrated to the center, which indicates that it is not affected by the X-ray heating from the ring collision, which dominates the H$\alpha$ morphology \citep{Larsson2011}.

\section{Origin of the H$_2$ emission}
\subsection{Spectral modeling}
\label{sec-mod}
To check that the identification of the lines with H$_2$ is consistent with wavelengths and expected relative fluxes, as well as the presence of other lines in this spectral window,  we have calculated synthetic spectra for different assumptions for the H$_2$ excitation.
 Because the temperature in the H rich gas in the SN core is $\la 200$ K \citep[][]{Jerkstrand2011,Kamenetzky2013}  thermal excitations of the vibrational levels are likely to be unimportant.  The remaining possibilities are excitation by  non-thermal electrons  or alternatively by UV fluorescence. Both these processes can lead to ionization or photodissociation of the H$_2$. Reformation of the molecule will then lead to vibrational-rotational excitation. \cite{Black1976}, however, argue that fluorescence should be more efficient, unless the UV flux is heavily absorbed. 
  
There are two sources of non-thermal  electrons in the ejecta. The ${}^{44}$Ti decay,  which dominates the radioactive input \citep{Jerkstrand2011,Boggs2015}, produces positrons, which results in a cascade of non-thermal electrons with energies $\ga 10$ eV. Whether  the positrons will reach the H rich blobs of the ejecta depends on the magnetic field.  Coulomb collisions alone cannot trap them in the iron rich regions  where the positrons are created \citep{Jerkstrand2011}, but even a weak field decreases the mean free path dramatically. It is, however, conceivable that  a fraction of the positrons may escape the ${}^{44}$Ti  sites and  may then ionize and excite the H rich parts of the core. 
The other source of fast electrons may be X-rays from the interaction with the ring. The evolution of the optical flux  from the ejecta   is now
dominated by these X-rays  \citep{Larsson2011}. When absorbed these give rise to fast electrons, and a similar cascade of non-thermal electrons as the positrons \citep{Fransson2013}.  A similar scenario has been advocated by \cite{Richardson2013} for the H$_2$ emission in the Crab nebula. The presence of a pulsar generating the relativistic particles makes, however, the situation different from the ejecta in SN 1987A.

\begin{figure*}[!t]
\begin{center}
 \resizebox{15.cm}{!}{\includegraphics[angle=-0]{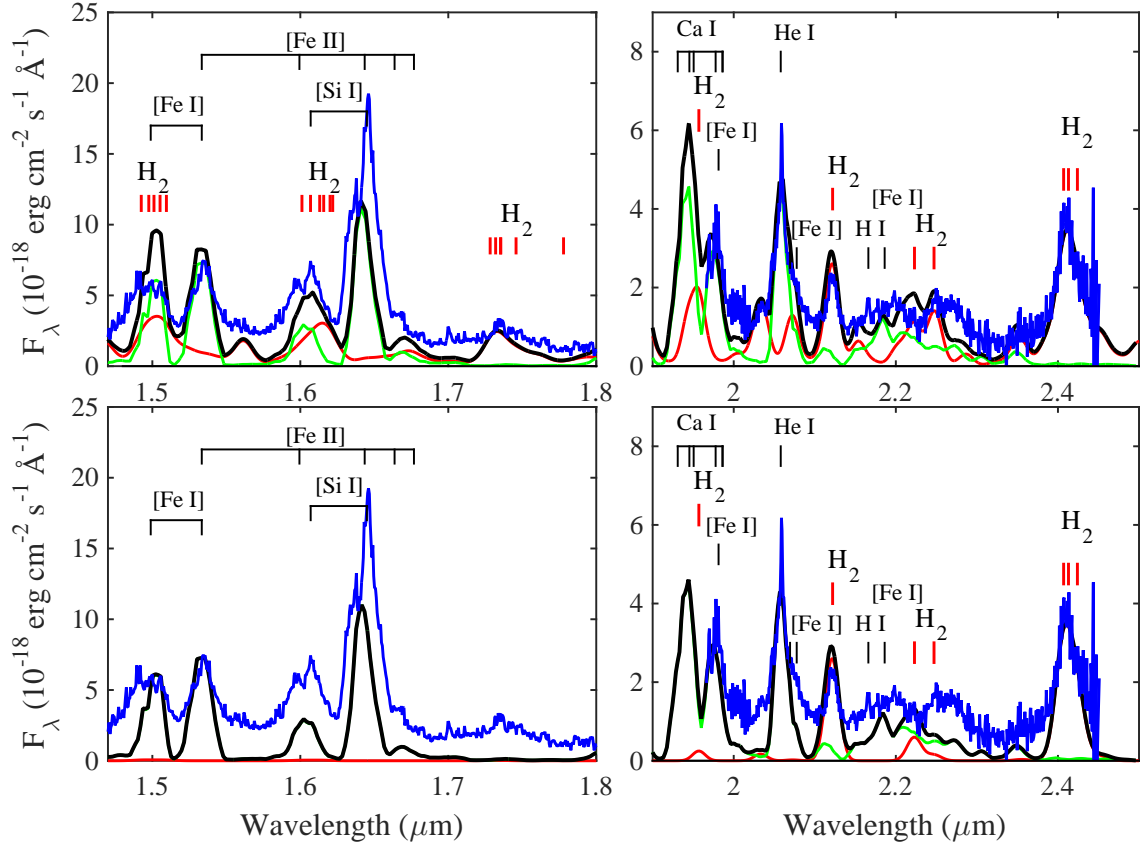}}
\caption{Upper panel: Coadded H and K band spectra from 2005 and 2007 (blue), together with the synthetic spectrum (black) for the UV dominated H$_2$ excitation model added to the ejecta spectrum from \cite{Kjaer2010}.   
The individual contributions from  H$_2$ (red), and the Kj{\ae}r et. al. spectrum (green)  are shown. Lower panel: Same for the non-thermal electron excitation model.  Note the absence of H$_2$ lines in the H band for this model.}
\label{fig3} 
\end{center}
\end{figure*}

 To test the non-thermal electron scenario we use the relative fluxes calculated  by \citet[][their Table 4]{Gredel1995}.  Note that the relative strengths of the lines depend  on the rotational temperature of the ground state levels, as well as the degree of ionization. The calculation referred to assumes a temperature of 300 K for the ground state populations and an electron energy of 30 eV. We have then normalized the line fluxes to that of the 2.122$\my$m line. 
 To approximately include the atomic lines we have also added the ${}^{44}$Ti powered ejecta spectrum from \cite{Kjaer2010}, scaled to the extraction aperture we use  (Fig. \ref{fig3}).

There may also be a weak continuum contribution.  
Synchrotron emission from the ring is observed at radio wavelengths with a spectrum $F_\nu \approx 25 (\nu/100 {\rm  \ GHz})^{-0.8}$ mJy \citep{Indebetouw2014}. Extrapolating  to 2.2 $\mu$m gives a  flux of $\sim 4.8 \times 10^{-18} \  {\rm erg \ s^{-1} \ cm^{-2} \ \AA^{-1}}$.
A minor fraction from high latitude emission above the ring plane could therefore contribute to the NIR continuum, unless synchrotron cooling would steepen the spectrum.  Dust emission from ultra-small grains could also contribute. From ALMA observations  we know there is a strong source of dust emission  in the core of the SN \citep{Matsuura2015} with  a temperature of $\sim 24$ K. \cite{Sarangi2015} find that the size distribution of grains  ranges from 
$\sim 10$ \AA \  to micron sized grains. 
 Ultra-small grains will
 be transiently heated by UV photons to very high temperatures \citep[e.g.,][]{Draine2011} and emit their radiation in the mid- to NIR. 
 While the absolute line fluxes depend on the continuum level, our qualitative results for the spectral models are not sensitive to these assumptions. Because of the uncertainty we have not included this component. 
 
 The line profiles are assumed to be Gaussian with FWHM $= 2300 \kms$,  the same as that found for the CO lines \citep{Kamenetzky2013}. We concentrate here on the co-added H- and K-band spectra from 2005 and 2007 which have the best S/N . 
 In the lower panel of
 Fig. \ref{fig3} we show these spectra together with the synthetic spectrum. 

UV fluorescence  from the H$_2$ ground state requires photons between 912 \AA \ and 1108 \AA, which may be created internally in the ejecta (Sect. \ref{sec-disc}), or alternatively from the ring collision.   
Ly$\alpha$ is strong from the ring \citep{France2011} 
and resonance fluorescence from the  H$_2$ $v=2$  levels  \citep{Shull1978} may be possible. This process, however,  requires either a temperature of $\ga 1000$ K or a sufficiently strong UV field for a significant population of this level. Between 912 \AA \ and $\sim 1150$ \AA \  little is known about the flux from the ring. 

For the UV fluorescence model we have taken the relative H$_2$ line strengths from Model 14 of \cite{Black1987} and again scaled this to the 2.122$\my$m flux.
The relative line strengths are not very sensitive to the specific shape of the spectrum in the far-UV or to the density as long as this is $\la 10^4 \ccm$. 

Comparing the two models, the relative flux of the 2.122$\my$m and the 2.40$\my$m lines is similar  and close to the observed ratio, especially considering the uncertain continuum level. This, together with the good fit of the line profiles,  confirms the H$_2$ line identifications. Note also the absence of atomic lines at both these wavelengths.

The main difference  between the two
models  is  the near absence of excitations to $v \geq 2$ in the non-thermal model. In the K-band \cite{Gredel1995} find the the (2,1)S(1) 2.248$\my$m/(1,0)S(1) 2.122$\my$m ratio to be 0.54  the UV dominated case, compared to 0.06 in the non-thermal electron case.  From Fig. \ref{fig3}  we see that the 2.248$\my$m and 2.223$\my$m lines give a  better fit to the feature at $\sim 2.25 \my$m for the UV fluorescence model. According to this model weaker lines at 1.953, 2.034, 2.071, 2.351$\my$m  may also be present in the observations. 
 Although the H-band is more crowded, we note that  there is in the UV model a line feature at 1.74$\my$m which does not coincide with any line included in the atomic model and coincides with a blend of comparatively strong H$_2$ lines between $1.73-1.75 \my$m. 
Also H$_2$ lines at 1.50 and  1.62 $\mu$m may possibly be present. 
The models in Fig. \ref{fig3} therefore favour the UV excitation model. This conclusion is, however, fairly marginal, given the S/N level of the spectrum  and uncertainties in both the H$_2$ and atomic line fluxes.  

From the line fits we estimate the total luminosity in the 2014 (day 10,120) spectrum of the 2.40$\my$m H$_2$ lines to $\sim 6.2 \times 10^{32} \ergs$ and of the 2.122$\my$m line to $\sim 2.4 \times 10^{32} \ergs$. This should be compared to the total input from positrons which is $\sim 10^{36} \ergs$, or  $\sim 1/3$ of the bolometric luminosity \citep{Larsson2011}. 
For the non-thermal electron case \cite{Gredel1995} finds that $ 46 \%$ of the H$_2$ emission is in the 1.95 - 2.45 $\mu$m range, while in the UV fluorescence case  15 \% is in this range. The total energy required for the IR H$_2$ lines is therefore $\sim 2 \times 10^{33} \ergs$ and $\sim 6 \times 10^{33} \ergs$, respectively.

\subsection{Time evolution}
\label{sec-tevol}

A distinguishing factor between the two different sources of energy for the H$_2$ emission, either  from the radioactive decay of ${}^{44}$Ti in the ejecta or from an external UV or X-ray flux from the ring collision, is the time evolution.   
From day 6500 to 9200 both the soft and hard   X-ray flux from the ring increased by a factor of $\sim 4.4$ \citep{Helder2013}. If the excitation is connected to the  external X-ray or UV flux we  expect the H$_2$ emission to increase by a substantial factor, while the energy input by ${}^{44}$Ti is expected to decrease slowly. 

A complication when measuring the flux evolution is the strong and wavelength dependent `continuum' level, which most likely is a blend of weak lines and true  continuum (e.g., Fig. \ref{fig3}). This makes absolute measurements of the fluxes  difficult and prone to systematic errors. Using the same prescription for the continuum level the trends should, however, not be affected. For the 'continuum' we use the average flux between $2.09-2.11 \ \mu$m for the \ion{He}{1} and the H$_2$ 2.122 $\mu$m lines and the $2.30-2.35\ \mu$m range for the H$_2$ 2.40 $\mu$m blend, which are chosen to be reasonably free of other emission lines. To minimize any remaining contribution from  scattered emission from the ring to the \ion{He}{1} line we  exclude for this line the central part within $\pm 700 \kms$ in the flux measurement. 

  From Table \ref{tab1} we conclude that the H$_2$ 2.40 $\mu$m luminosity is nearly constant within the statistical errors, which we estimate to $0.15\times 10^{32} \ergs$, while the \ion{He}{1}  line shows an increase by a factor of $\sim 2.3$ from day 6832 to 10,120. The low  H$_2$ 2.122 $\mu$m flux makes it  difficult to draw any firm conclusion from this, although it is consistent with a constant flux within the errors. Note  that these luminosities are likely to be systematically underestimated by a considerable factor, as is e.g., seen from the models in Fig. \ref{fig3}. The absolute luminosities given in Sect. \ref{sec-res} are probably more accurate. Here we are more concerned with the time evolution, without making  assumptions about the excitation mechanism. 
 \begin{deluxetable}{rccc}
\tablecolumns{4}
\tablewidth{0pc}
\tablecaption{H$_2$ and \ion{He}{1} luminosities\tablenotemark{a}. See text for errors. \label{tab1}}
\tablehead{
\colhead{Epoch} & \colhead{H$_2$ 2.122\ $\mu$m}   & \colhead{H$_2$ 2.40 \ $\mu$m}& \colhead{\ion{He}{1} 2.058 \ $\mu$m}}
\startdata
6832&0.37&3.11&1.10\\
 7606&   0.45   &2.88&   1.35\\
8694&   0.57  &3.32& 1.59\\
10,120 &   0.64 &3.45&   2.54\\
\enddata
\tablenotetext{a}{In units of $10^{32} \ergs$}
\end{deluxetable}
 
The nearly constant H$_2$ luminosities indicate that  these lines are powered by the  ${}^{44}$Ti decay, either through the UV generated in the ejecta or possibly through leaking positrons. Gamma-rays from the ${}^{44}$Ti decay gives an input decreasing with time as $\tau_{\gamma} \propto 1/t^2$. They, however,  only contribute marginally to the H$_2$ luminosity.

\section{Discussion and conclusions}
\label{sec-disc}

Summarizing, we find that the spectrum marginally favours the UV fluorescence model and that it is powered by the ${}^{44}$Ti decay. These may together be realized if the excitation is dominated by  diffuse internal UV radiation in the ejecta. 

The H$_2$ emission is  concentrated to the core, as is the case for the [\ion{Si}{1}]/[\ion{Fe}{2}]  1.644$\my$m line (Fig. \ref{fig2}). This result is consistent with the absence of time evolution. The presence of H$_2$ in the core is also consistent with the H$\alpha$ profile 
from early observations,  which was seen down to $\la 700 \kms$ \citep{Kozma1998b}. 
Outside the core the UV and X-ray emission from the ring collision may dissociate the H$_2$, while  these X-rays are responsible for most of the H$\alpha$ emission, resulting in the 'horse-shoe' shape in Fig. \ref{fig2}. The core itself is largely shielded from X-rays by the high metallicity \citep{Fransson2013}. 
Also the UV  from the ring may be absorbed at energies higher than the \ion{C}{1} photoelectric threshold at 11.2 eV ($< 1100$ \AA), depending on the covering factor of the C rich regions relative to the H rich in the core. 

The UV excitation scenario requires an internally generated UV field, like the \ion{He}{1} two-photon continuum, as proposed by  CMC \citep[see also][]{Jerkstrand2011}. 
 CMC also point out that the total opacity of the H rich region in the range $912 < \wl \la 1400$ \AA \ should be  dominated by resonance scattering in the Lyman and Werner bands of H$_2$, converting the radiation in this region into fluorescence lines between 1400 \AA \ and $\sim 1700$ \AA. 
\cite{Sternberg1989b} has calculated this in  detail, based on an ISM-like UV spectrum. Of particular interest is the strong blend of lines at 1590-1610 \AA, which also stands out in the simulations of  CMC. In the COS spectrum by \cite{France2011} there is indeed a weak feature at this wavelength. The spectrum is, however, complex and the significance of this feature is marginal.

Most of the H$_2$ formation in 
the CMC models took place before $\sim 1000$ days after explosion. 
NIR observations between days 377 to 1114 \citep{Meikle1993} showed a weak feature which they proposed could be due to the 2.122$\my$m line., but noted that this could be due to [\ion{Fe}{2}] 2.133 $\my$m.  In the spectra before 600 days this feature was much stronger than in the models by CMC, while at later epochs it was consistent with the models.  The observations of \cite{Meikle1993} did not show the 2.40$\my$m line, so no definite conclusion could be drawn. 

Other H$_2$ lines  expected in the UV case include blends  at $\sim$ 1.24, 1.31, and  1.40 $\mu$m. The J-band is, however,  more crowded than the K-band and has a lower S/N and Strehl ratio and is therefore here not shown. The lowest excitation lines of H$_2$ at 28.211$\my$m and 17.030$\my$m are expected to have fluxes $á\sim 1.5$ and $\sim 1.0$ times  the 2.122$\my$m line, respectively. 

Finally, the fact that the \ion{He}{1}  line  shows an  increase similar to H$\alpha$ \citep{Fransson2013} implies that at least part of this emission should  be affected by the X-ray input from the ring collision. This is somewhat surprising, given the similar distribution of this emission to that of H$_2$  (Fig. \ref{fig2}).
The \ion{He}{1}  emission, however, has contributions both from the H, He and Fe/He zones \citep{Kjaer2010}. The Fe/He component, together with the mixed-in H- and He-rich material in the core, may therefore explain the centrally peaked \ion{He}{1} emission,  while the former two components in the envelope may be responsible for the luminosity increase, which is mainly seen in the line wings, as for H$\alpha$ \citep{Larsson2015}. 

In conclusion, the discovery of H$_2$ in SN 1987A adds a new and important diagnostic of the conditions in the ejecta and may, together with observations with ALMA, provide new insight into the chemistry of the H rich regions of the core \citep[e.g.,][]{Cherchneff2009}. 
\acknowledgments
This research was supported by the Swedish Research Council and the Swedish
National Space Board. Based on observations collected at the European Organisation for Astronomical Research in the Southern Hemisphere.

\end{document}